\documentclass[12pt]{article}
\usepackage{epsfig}
\usepackage{amssymb}
\usepackage{amsmath}

\newcommand{\za}{{\alpha}}   
\newcommand{\zb}{{\beta}}    

\newcommand{\ZA}{{A}}   
\newcommand{\ZB}{{B}}   

\newcommand{\p}{{+}}

\newcommand{\n}{{-}}
\newcommand{\yp}{{y_{+}}}
\newcommand{\yn}{{y_{-}}}

\newcommand{\ddim}{{d}}
\newcommand{\DDim}{{(d+\!1)}}

\newcommand{\M}{{\cal M}}
\newcommand{\dM}{{\partial \cal M}}
\newcommand{\dx}{{d^{^{\,\ddim}} \!x}}

\newcommand{\I}{{\mathbf{I}}}
\newcommand{\un}{\mathbf n} 

\newcommand{\ddy}{{\partial_y}}

\newcommand{\htt}{{h}}   

\newcommand{\bBox}{\square}  
\newcommand{\BBox}{\hat{\mathbf{F}}}  

\newcommand{\bnabla}{\bigtriangledown}  
\newcommand{\Bnabla}{\nabla}  
\newcommand{\Bnablan}{\nabla_{\un}}

\newcommand{\GrN}{ \mathbf{G}_{N}}  
\newcommand{\GrD}{ \mathbf{G}_{D}}  

\newcommand{\nGrDn}{\overrightarrow{\mathrm{W}}\mathbf{G}_{D}\overleftarrow{\mathrm{W}}} 

\newcommand{\NGrDN}{\overrightarrow{\mathrm{W}}\mathbf{G}_{D}\overleftarrow{\mathrm{W}}} 

\newcommand{\RnGrDnR}{{\NGrDN}} 
\newcommand{\rnGrDnr}{{\nGrDn}} 

\newcommand{\D}{\mathbf{D}}  
\newcommand{\E}{\mathbf{E}}  
\newcommand{\C}{\mathbf{C}}  

\newcommand{\tens}{{\sigma}}

\newcommand{\R}{{\mathbb R}} 
\newcommand{\dV}{dV\,}
\newcommand{\dS}{dS}

\begin{document}

\title{\bf Duality of boundary value problems and braneworld action
in curved brane models}
\author{~A.~O.~Barvinsky$^{\dag}$ and  ~D.~V.~Nesterov$^{\star}$}
\maketitle
 \begin{center}

  \hspace{-0mm}{\em Theory Department, Lebedev Physics
  Institute,
  Leninsky Prospect 53, Moscow 11999,\\ Russia}
 \end{center}

\begin{abstract}
Braneworld effective action is constructed by two different
methods based respectively on the Dirichlet and Neumann boundary
value problems. The equivalence of these methods is shown due to
nontrivial duality relations between special boundary operators of
these two problems. Previously known braneworld action algorithms
in two-brane Randall-Sundrum model are generalized to curved
branes with deSitter and Anti-deSitter geometries.
\end{abstract}
\noindent
$^{\dag}$e-mail: barvin@lpi.ru\\
\noindent
$^{\star}$e-mail: nesterov@lpi.ru\\

\section{Introduction}
\hspace{\parindent}Rapidly developing theory of braneworld
phenomena \cite{Ant,string,hierarchy} requires new efficient
methods of their description. These methods incorporate together
with the well-known old formalisms, like the effective action
approach, special new features associated with bulk/brane
(boundary) ingredients characteristic of the braneworld scenarios.
In this paper we want to focus our attention at the peculiarities
of the Dirichlet and Neumann boundary conditions in braneworld
setup, the way they arise in the course of calculating braneworld
effective action and, in particular, at a sort of duality relation
between these problems. This relation manifests itself in the
equality of special boundary operators originating from Dirichlet
and Neumann boundary value problems and serving as a kernel of the
braneworld effective action in the approximation quadratic in
fields.

Braneworld effective action implicitly incorporates the dynamics
of the fields in the bulk and explicitly features the boundary
fields (being the functional of those) which, in braneworld
scheme, are directly observable by the observer living on the
brane. Such action, on the one hand, arises as a result of
integrating out the bulk fields subject to boundary (brane) fields
and, on the other hand, generates effective equations of motion
for the latter. This situation obviously suggests two strategies
of calculating the braneworld action. One strategy consists in its
direct calculation and, in the tree-level approximation, reduces
to solving the equations of motion in the bulk subject to given
boundary values -- brane fields -- and substituting the result
into the fundamental bulk action. Thus, this strategy involves the
Dirichlet problem. Another strategy consists in recovering the
braneworld effective action from effective equations of motion for
brane fields. Since the latter incorporate well known Israel
junction conditions on branes \cite{Israel}, this method relies on
the Neumann boundary value problem. The consistency of these two
methods, which obviously should lead to one and the same result,
is far from being explicit. Here we show that they really match in
view of a nontrivial relation between nonlocal brane operators
arising as restriction of (properly differentiated) kernels of
Dirichlet and Neumann Green's functions to the boundaries/branes.
As a byproduct of the Dirichlet setup we derive the answer for the
braneworld effective action in models with curved branes of
deSitter and Anti-deSitter geometries.

Curved brane models have been considered in many papers their very
incomplete list consisting of
\cite{GarSas,Reall,HHR,ParikhSol,GenSas,GarPujTan}. In contrast to
the results of these works we concentrate on two-brane models with
{\em independent} metric configurations on their branes. This is
done along the lines of the previous papers \cite{brane,BWEA}
where the weak field (quadratic in perturbations) approximation
for braneworld action was obtained on the background of flat
branes. In distinction from \cite{HHR,GenSas,GarPujTan} we obtain
the nonlocal action nontrivially intertwining metric fields on
both branes. The interest in such a construction follows from the
attempts of describing off-shell phenomena in braneworld physics,
solving the hierarchy and cosmological constant problems,
generating inflation in braneworld scenarios \cite{Tye,brane},
etc.

The paper is organized as follows. In Sect.2 we define the
braneworld effective action and sketch its calculation within the
perturbation expansion with Dirichlet boundary conditions on
branes. In Sect.3 we apply this technique to the two-brane
Randall-Sundrum model \cite{RS} to find its action in the
approximation quadratic in the perturbations of metric fields on
branes. This result generalizes the algorithms of the works
\cite{brane, BWEA} to curved branes of deSitter and Anti-deSitter
geometries. Its derivation is based on the technique of linearized
invariants of the diffeomorphism gauge transformations intensively
used in the theory of cosmological perturbations
\cite{BFM,GarMonTanSas,efeqmy}. Sect.4 gives a review of the
alternative calculation \cite{BWEA} of the same result, based on
its recovery from the effective equations of motion. In Sect.5 we
show the equality of the two manifestly different quadratic forms
of the action by proving that their kernels are equal to one
another -- the main duality relation between specific operators of
the Dirichlet and Neumann boundary value problems. Finally, in
concluding section we discuss possible implications of the
obtained results in braneworld scenarios with curved (inflating)
branes.

\section{Braneworld effective action}
\hspace{\parindent}We will calculate the braneworld effective
action defined by the procedure that was discussed in much detail
in \cite{BWEA}. In contrast to the conventional Kaluza-Klein
reduction, usually used for the construction of the effective
action \cite{KubVol,Kub}, our action is a functional of the fields
that have direct physical and geometrical interpretation. They
coincide with the restriction $\phi$ of the fundamental bulk
fields $\Phi$ to the branes, they are directly observable by
observers living on these branes, and serve as boundary conditions
for $\Phi$.

To be more precise, let $\Phi$ be a set of fundamental dynamical
fields on $(d+1)$-dimensional spacetime $\M$ (bulk) with the full
boundary $\dM$. In $\dM$ we shall include not only the ``outer''
boundary (like asymptotic domains of the bulk spacetime) but also
the branes -- timelike $d$-dimensional surfaces embedded into $\M$
and enumerated by the index $i$, $\dM=\bigcup_{i}\dM_i$. Boundary
values of $\Phi$ on $\dM_i$ we denote by $\phi_i$,
$\Phi|_{_{\dM_i}}=\phi_i$, and call them brane (or induced)
fields. In addition to the bulk $(d+1)$-fields there are usually
$d$-dimensional fields $\varphi$ which, by definition, live only
on branes and which will be in what follows denoted as matter
fields\footnote{Usually the role of bulk fields is played by the
bulk spacetime metric, because string theory motivated braneworld
framework does not admit the propagation of gauge (matter) fields
in the bulk \cite{string}.}.

Thus, the full $(d+1)$-dimensional action is
    \begin{eqnarray}
     \mathbf{S}[\,\Phi,\varphi\,]=
     \mathbf{S}^{\DDim}[\,{\Phi}\,]
     +S^{\mathrm{mat}}[\,\phi,\varphi\,],       \label{FAction}
    \end{eqnarray}
where the first term denotes the bulk part depending only on
$\Phi$, while the second term represents the matter action which
depends only on induced (brane) $\phi$ and matter fields
$\varphi$. For brevity we omit the index $i$ of brane and matter
fields, so that $(\phi,\varphi)$ denotes the collection of fields
$(\phi_i,\varphi_i)$ associated with all the branes.

Braneworld effective action is the result of integrating out the
bulk fields subject to given values of brane fields
    \begin{eqnarray}
    \exp\Big(iS^{\mathrm{eff}}[\,\phi,\varphi\,]\Big)
    =\left.\int D\Phi\,
    \exp\Big(i\mathbf{S}[\,\Phi,\varphi\,]\Big)
    \right|_{\Phi(\dM)=\phi}.                 \label{BdefBWEA}
    \end{eqnarray}
Since the action of matter fields enters (\ref{FAction})
additively and its arguments are not integrated over, it continues
entering $S^{\mathrm{eff}}[\,\phi,\varphi\,]$. The rest of the
latter is highly nontrivial, because it accumulates the result of
the functional integration. When this integration is done within
$\hbar$-expansion the result reads
    \begin{eqnarray}
     S^{\mathrm{eff}}[\,\phi,\varphi\,]
     =S_d[\,\phi\,]
     +S^{\mathrm{loop}}[\,\phi\,]
     +S^{\mathrm{mat}}[\,\phi,\varphi\,],  \label{DefBWEALoopExp}
    \end{eqnarray}
where the tree-level part
    \begin{eqnarray}
     S_d[\,\phi\,]=
     \mathbf{S}^{\DDim}\big[\,\Phi[\,\phi\,]\,\big]     \label{S}
    \end{eqnarray}
is a result of substituting in the classical bulk action the
solution $\Phi[\,\phi\,]$ of the following  Dirichlet boundary
value problem
    \begin{eqnarray}
    \left\{ \begin{array}{l}
    \displaystyle{\frac{\delta\mathbf{S}^{\DDim}}
    {\delta \Phi}}=0,  \\
    \\
    \Phi|_{_{\dM}}=\phi.
     \end{array} \right.                        \label{F(f)}
    \end{eqnarray}
Also, $S^{\mathrm{loop}}[\phi]$ represents the loop part, of order
$\hbar$, which we shall disregard in what follows.

When considering the full nonlinear theory (\ref{FAction}) it is
generically impossible to obtain the tree-level term (\ref{S}) as
an exact explicit functional of induced fields $\phi$. Instead,
one can develop a perturbation theory for (\ref{S}) in powers of
the perturbation of boundary conditions
    \begin{eqnarray}
    \phi=\phi^0+h                   \label{h}
    \end{eqnarray}
on the background of some configuration $\phi^0$ at which the
effective action is exactly calculable (or represents an
irrelevant constant). This perturbation induces the perturbation
$H[\,h\,]$ of the solution of the boundary problem in the
bulk (\ref{F(f)})
    \begin{eqnarray}
    \Phi[\,\phi\,]=
    \Phi^0+H[\,h\,],         \label{DefFieldsExp}
    \end{eqnarray}
where $\Phi^0=\Phi[\,\phi^0]$ is the bulk background solving the
classical equations of motion with background boundary conditions
on branes $\phi=\phi^0$, which is also supposed to be explicitly
known.

From (\ref{F(f)}) and (\ref{DefFieldsExp}) it follows that up to
$O(h^2)$-terms, the bulk perturbation $H[\,\phi^0,h\,]$ satisfies
the following linear Dirichlet problem:
    \begin{eqnarray}
    \left\{ \begin{array}{l}
    \BBox H=0, \\
    \\
    H|_{_{\dM}}=h,
     \end{array} \right.                      \label{dF(df)}
    \end{eqnarray}
where $\BBox$ is the operator of small field disturbances in the
theory with the action (\ref{FAction}). Since we assume that its
Lagrangian contains at most first-order derivatives, $\BBox$ is a
second-order differential operator with the kernel in the space of
$(d+1)$-dimensional bulk coordinates $X$ calculated on the
background $\Phi^0$
    \begin{eqnarray}
    \BBox\delta(X,X')=
    \left.\frac{\delta^2 \mathbf{S}^{\DDim}[\,\Phi\,]}{\delta
    \Phi(X)\,\delta\Phi(X')}\,\right|_{\Phi=\Phi^0}.
    \end{eqnarray}

As any other second-order differential operator, $\BBox$ satisfies
the Wronskian relation following from integration by parts in the
equation
    \begin{eqnarray}
     \int\limits_{\M} \dV
    \left(\Phi_1(\BBox \Phi_2)
    -(\BBox\Phi_1) \Phi_2\right)=
     -\int\limits_{\dM}\dS\,
    \left(\Phi_1 (\hat W\Phi_2)
    -(\hat W \Phi_1)\Phi_2\right)         \label{Wronskian}
    \end{eqnarray}
    for arbitrary $\Phi$ and $\Psi$,
for arbitrary test functions $\Phi_1$ and $\Phi_2$. Here
$\dV$ is the bulk integration measure and $\dS$ is the measure on the
boundary, and we shall call $\hat W$ the {\em  Wronskian} operator. It is
of the first order in derivatives and, for regular situations, it
necessarily contains the derivative normal to the boundary. Obviously,
this operator is defined by the relation
(\ref{Wronskian}) only up to addition of an arbitrary selfadjoint operator
on the boundary. It can be specified uniquely by demanding that
    \begin{eqnarray}
    \int\limits_{\M} \dV \Phi_1
    \overleftrightarrow{\mathbf F}\Phi_2  =
    \int\limits_{\M} \dV \Phi_1(\BBox \Phi_2)+
        \int\limits_{\dM}\dS\,
    \Phi_1(\hat W \Phi_2),        \label{Wronskian2}
    \end{eqnarray}
where the notation $\overleftrightarrow{\mathbf F}$ implies that
the derivatives of $\BBox$ are acting (in the sense of integration
by parts) so that the left-hand side represents the form bilinear
in first order derivatives. For $\Phi_2=\Phi_1$ the integrand of
the left-hand side is just the Lagrangian quadratic in the field
and its first-order derivatives. In this case the operator $\hat
W$ can also be obtained in terms of the canonical momentum
conjugated to the field relative to ``time'' normally flowing to
the boundary \cite{reduc}.

This Wronskian operator allows one to formulate in
a closed form the solution of the Dirichlet problem
(\ref{dF(df)}) given the Green's function of this problem
satisfying
    \begin{eqnarray}
     \left\{ \begin{array}{l}
     \BBox \GrD(X,X')=\delta(X,X'), \\
    \\
     \GrD(X,X')|_{_{X\in\dM}}=0.
     \end{array} \right.                   \label{DGfbvp}
    \end{eqnarray}
This solution reads as the following integral over the boundary
$\dM$
    \begin{eqnarray}
        \Phi(X)=-\int \limits_{\dM}\dS'
        \Big[\,\GrD(X,X')
    \overleftarrow{W'}\,\Big]_{X'=X(x')}
    h(x').                        \label{DGenSol}
    \end{eqnarray}
Here $X=X(x)$ denotes the embedding of the boundary $\dM$
parameterized by internal coordinates $x$ into the bulk with
coordinates $X$, $dS'=dS(x')$ is the surface integration element
at $x'$ and the arrow over $W'$ indicates that the derivatives of
the primed Wronskian operator act to the left on the second
(primed) argument of the Green's function.

The expansion of the action in the field perturbation (\ref{DefFieldsExp})
begins with
    \begin{eqnarray}
     S_d[\,\phi\,]
    =\mathbf{S}^\DDim\big[\,\Phi^0\big]
     +\frac12\,\int\limits_{\M} dV\,
    H\overleftrightarrow{\mathbf F}H
    +O[h^3],                          \label{DefBWEAPertExp}
    \end{eqnarray}
because the linear term identically vanishes in virtue of
classical equations of motion for $\Phi^0$ and, what is important,
in virtue of the Israel matching conditions on branes, which
annihilate the linear surface terms on boundaries even despite the
fact that $H|_{\dM}=h\neq 0$. Note that the quadratic part of this
expansion is built with the aid of the operator
$\stackrel{\leftrightarrow}{\mathbf F}$ symmetrically acting on
both fields -- the consequence of the fact that the initial action
did not contain second-order derivatives. Therefore, using the
relation (\ref{Wronskian2}) and taking into account linear
equations of motion for perturbations (\ref{dF(df)}) one finds
that the bulk part of the quadratic term vanishes and the
quadratic approximation for the braneworld effective action
reduces to the surface term\footnote{In what follows we restrict
ourselves with the quadratic approximation and omit terms
$O[h^3]$.}
    \begin{eqnarray}
     S_{\ddim}[\,\phi\,]
     =\mathbf{S}^\DDim\big[\,\Phi^0\big]
     +\frac12\,\int\limits_{\dM} \dS\,
     H(\hat W H).          \label{DefBWEAPertExp2}
    \end{eqnarray}
Further, a substitution of (\ref{DGenSol}) in this expression
gives the general answer for the quadratic action
    \begin{eqnarray}
     S_{\ddim}^{(2)}[\htt]= -\frac12 \int
     \limits_{\dM}\dS\int\limits_{\dM}\dS'
     h(x){\RnGrDnR}(x,x')\,
     h(x'),                        \label{DActionFinal0}
    \end{eqnarray}
where the kernel of the integral operation on the boundary
${\RnGrDnR}(x,x')$ is given by an obvious sequence of operations with the
kernel of the Dirichlet Green's function -- acting on it from both sides
with Wronskian operators and restricting the result to the boundary
with respect to both arguments
    \begin{eqnarray}
     {\RnGrDnR}(x,x')=\Big[\,\overrightarrow{W}\GrD(X,X')
     \overleftarrow{W'}\,\Big]_{X=X(x),\,X'=X(x')}.
    \end{eqnarray}

As mentioned above, for the case of several branes or spacetime
boundary consisting of several pieces $\dM=\bigcup_i\dM_i$ the
boundary (brane) fields denote the collection $\phi=\phi^i$.
Correspondingly the brane embedding functions and respective
perturbations acquire the same index, $X=X_i(x)$, $h=h^i\equiv
h|_{\dM_i}$, and the equations above imply the following obvious
generalization
    \begin{eqnarray}
     &&\int\limits_{\dM}\dS\to
     \sum\limits_i\int\limits_{\dM_i}\dS,\\
     &&{\RnGrDnR}(x,x')\to
     [{\RnGrDnR}]_{ij}(x,x')=
     \Big[\,\overrightarrow{W}\GrD(X,X')
     \overleftarrow{W'}\,\Big]
     _{X=X_i(x),\,X'=X_j(x')}.         \label{matrix}
    \end{eqnarray}
Under these replacements the expression (\ref{DActionFinal0}) becomes
a quadratic form with the nonlocal matrix-valued kernel (\ref{matrix}).

\section{Two-brane Randall-Sundrum model}
\label{Dirichlet}
\renewcommand{\M}{\mathbf B}
\renewcommand{\dM}{\mathbf b}
\hspace{\parindent}Here we specify the construction of the above
type on the example of the two-brane Randall-Sundrum model
\cite{RS} with the action
    \begin{eqnarray}
     &&\mathbf{S}[\,G,\varphi\,]=\int\limits_{\M\times\mathrm{Z}_2}
     d^{d+1}X \sqrt{G}\,
     \big(R(G)-2\Lambda\big)-
     2\sum_{i}\int\limits_{\dM_i}
     \dx\sqrt{g}\,[K] -\sum_i
     \int\limits_{\dM_i}
     \dx\sqrt{g}\, \tens_i+ \nonumber \\
     &&\qquad\qquad\qquad\qquad
     +\sum_i\int\limits_{\dM_i}\dx\sqrt{g}\,
     L^{\mathrm{mat}}
    (g,\varphi,\partial\varphi).         \label{Action}
    \end{eqnarray}
The orbifold symmetry implies that the $(d+1)$-dimensional
integration runs over two identical copies of the bulk $\M$ which
is bounded by two $\ddim$-dimensional branes $\dM_i$, $i=\pm$,
that can be regarded as the boundaries of $\M$. Here $G=G_{AB}(X)$
($A=0,1,...,d$) is the bulk metric and $g=g^\pm_{\alpha\beta}(x)$
($\alpha=0,1,...,d-1$) denotes the collection of induced metrics
on $i=\pm$ branes, $[K]$ is a jump of the trace of the extrinsic
curvature on the brane defined as $K= G^{\ZA\ZB}\nabla_\ZA
\un_\ZB$, where $\Bnabla_\ZA$ is a covariant $\DDim$-dimensional
derivative and $\un$ is an inward pointing normal. With this
definition the normals on two sides of the brane are oppositely
oriented and the extrinsic curvature jump $[K_{\alpha\beta}]$
actually equals the sum of the so-defined curvatures on both sides
of the brane. The orbifold symmetry obviously implies
$\int_{\M\times\mathrm{Z}_2}(...)=2\int_\M(...)$ and $[K]=2K$.
Finally, $\Lambda$ is the bulk cosmological constant and $\tens_i$
are brane tensions.

Our goal is to construct the perturbative braneworld effective
action for such a system. The fundamental bulk field is the bulk
metric $\Phi=G_{\ZA\ZB}(X)$, the role of brane field is played by
$\phi=g^\pm_{\za\zb}(x)$ -- the induced metric on branes. The
first three terms of (\ref{Action}) comprise the bulk part of the
action and the last term is the action of matter fields located on
branes\footnote{For reasons of convenience we include the brane
tension terms into the bulk part of the action. As far as it
concerns the Gibbons-Hawking term with the trace of extrinsic
curvature, it should also be related to the bulk part because it
makes by integration by parts the whole Einstein-Hilbert action
quadratic in first-order derivatives.}.

We choose as the background $\Phi^0$ the well-known
Randall-Sundrum solution in this model. In the setting generalized
to the case of curved branes this solution arises as follows.
Assume that there exists a coordinate system on the
$(d+1)$-dimensional bulk $X^\ZA=(x^\za,y)$ in which the background
metric solving the vacuum Einstein equations has the form
    \begin{eqnarray}
     G^0_{\ZA\ZB}(X)= \left(\begin{array}{cc}
     a^2(y)g_{\za\zb}(x) & 0\\
     0 & 1 \end{array}\right).        \label{Background}
    \end{eqnarray}
Two branes bounding the bulk manifold $\M$ are the hypersurfaces of
constant $y$ located at $y=y_\p$ and $y=y_\n$, so that $\M$
has a topology of $\dM\!\times\!\I$ with $x^\za$ -- the
coordinates on $\ddim$-dimensional $\dM$ and $\I$ -- the
interval of the coordinate $y\in[y_\p,y_\n]$. In the wording of
the previous section the boundary conditions $\phi^0$ for this
bulk background $\Phi[\phi^0]$ are two induced metrics conformally
equivalent to one another $g^\pm_{\za\zb}=a^2(y_\pm)g_{\za\zb}(x)$
and conformally related to some fixed metric $g_{\alpha\beta}(x)$. With
the ansatz (\ref{Background}) this metric is restricted by the
condition of constant scalar curvature. Indeed, from the $yy$-component
of the vacuum Einstein equations in the bulk it follows that
    \begin{equation}
     a^{-2}(y)R(g)-2\Lambda =
     \ddim(\ddim-1)\left(\ddy
     \ln a(y)\,\right)^{2},      \label{EquationFora}
    \end{equation}
where $R(g)$ is the scalar curvature of $g_{\alpha\beta}(x)$.
Therefore, only metrics with constant $R(g)$ independent of $x$
can satisfy this equation. They include, in particular, three
maximally symmetric cases -- deSitter ($dS^d$), plane
($\R^{\ddim-1,1}$) and Anti-deSitter ($AdS^{\ddim}$) branes
embedded into the Einstein space with the cosmological constant
$\Lambda$. As in the Randall-Sundrum case we shall consider the
case of anti-deSitter bulk with negative $\Lambda$
    \begin{eqnarray}
     \Lambda= -\frac12 \ddim(\ddim-1)k^2,
    \end{eqnarray}
so that Eq.(\ref{EquationFora}) gives the following profiles of
the scale factor $a(y)$ for the foliation of the Anti-deSitter
spacetime by the deSitter, flat and Anti-deSitter slices
    \begin{eqnarray}
     a^2(y)=\left\{\begin{array}{ll}
     \displaystyle{\frac{\mathcal{H}^2}{k^2}}
     \sinh^2(ky),\quad &\; dS^\ddim,\\
     \exp(-2ky), \quad &\; \R^{\ddim-1,1},\\
     \displaystyle{\frac{\mathcal{H}^2}{k^2}}
     \cosh^2(ky), \quad &\; AdS^\ddim,
     \end{array} \right.                          \label{dSRAdS}
    \end{eqnarray}
where the ``Hubble'' constant $\mathcal{H}$ is given by
    \begin{eqnarray}
     \mathcal{H}^2=\frac{\big|R(g)\big|}{\ddim(\ddim-1)}.
    \end{eqnarray}

Among these three cases the original Randall-Sundrum model
\cite{RS} corresponds to the foliation with flat slices. Two flat
and empty branes coincide with the pair of such slices and serve
as boundaries of a smooth piece of Anti-deSitter bulk at $y_+\leq
y\leq y_-$. With the orbifold symmetry the piecewise smooth
Anti-deSitter solution of vacuum Einstein equations satisfies the
Israel junction conditions on flat branes when their tensions are
opposite in sign and equal\footnote{For simplicity we dropped in
Eq.(\ref{Action}) the gravitational coupling constant so that the
action has the dimensionality of length to the power $d-1$ and the
brane tensions have the dimensionality of inverse length.}
    \begin{equation}
     \tens_\p=-\tens_\n
     =4\sqrt{\frac{2(\ddim-1)
     |\Lambda|}{\ddim}}.\label{TensionsFlat}
    \end{equation}

Other cases of curved (deSitter and Anti-deSitter) branes
correspond to different expressions for their tensions
$\sigma_i=\sigma_\pm$. For $Z_2$-orbifold interpolating between
the two curved branes these tensions, when they satisfy Israel
junction conditions, depend on the location of branes $y_i=y_\pm$
and read
    \begin{equation}
     \tens_i=-4(\ddim-1)\,
     \Bnabla_{\un}\ln{a(y)}\Big|_{y_i},      \label{TensionsGen}
    \end{equation}
where we introduce the notation
    \begin{eqnarray}
     \Bnabla_{\un} \equiv\un^\ZA
     \partial_\ZA=
     \left\{\begin{array}{l}
     +\ddy, \qquad y=\yp,\\
     -\ddy, \qquad y=\yn,
     \end{array}\right.             \label{Def_nabla_n}
    \end{eqnarray}
for the {\em inward} pointing derivative normal to $\pm$-branes in
the background metric (\ref{Background}). In order to match with
the inward orientation of the normal vector (or location of the
positive tension brane to the left of the negative tension one) we
have to assume that the coordinate $y$ in (\ref{dSRAdS}) is
negative for $dS$ and $AdS$ cases, $y_\p < y_\n < 0$. This also
guarantees that deSitter branes are not separated by particle
horizon at which $a(y)=0$ \cite{ParikhSol,BWSolutions} and, thus,
bound a causally connected bulk. Conversely, relations
(\ref{TensionsGen}) can be viewed as the equations which for given
bulk curvature parameter $k$ and two different brane tensions of
opposite signs $\tens_\p,\,\tens_\n$ determine the locations of
branes at $y_\p$ and $y_\n$.

Our goal now is to build, along the lines of Sect.2, the
perturbation expansion for braneworld action on this background by
perturbing the brane configurations like in (\ref{h}) and studying
the response to it in the bulk (\ref{DefFieldsExp}). The
geometrical picture of the background solution above actually
shows that a simple scheme of Sect.2 should be supplemented by a
number of important modifications due to local diffeomorphism
invariance of the theory. This invariance manifests itself in many
facts. In particular, not all components of the bulk metric
$G_{AB}(x,y)$ should be fixed as boundary conditions on the
branes, but rather the induced metric coefficients
$g_{\alpha\beta}$, whose number is less than that of
$G_{AB}$\footnote{The distinguished status of the
$G_{yA}$-components of the metric that should {\em not} be fixed
at the boundary $y={\rm const}$ follows from the fact that they
serve as Lagrange multipliers of constraints in the 'Hamiltonian'
formalism associated with the $y$-`time' foliation of spacetime.
This in turn follows from the local gauge invariance of the
Einstein theory.}. Moreover, the induced metric coefficients are
nontrivially related to $G_{AB}(x,y)$ depending on the embedding
of boundaries in the bulk, this embedding being nontrivial after
the inclusion of gravitational perturbations and depending on the
gauge fixation of the latter.

These properties of the braneworld formalism were considered in
much detail in a number of papers \cite{BWEA,GarTan,ChGR} where
the location of branes in the bulk was described in the Gaussian
normal coordinates, $G_{yy}=1,\,G_{y\mu}=0$, by an additional
scalar field -- the radion. Here we prefer to proceed in a more
gauge invariant manner without explicitly fixing the coordinate
system in the bulk. The only restriction on the choice of
$(d+1)$-dimensional coordinates will be the requirement that the
branes are located at constant values of the
$y$-coordinate\footnote{At least for small metric perturbations
this is always possible, because the foliation by surfaces of
constant $y$, Eq.(\ref{dSRAdS}), is regular in patches of the
Anti-deSitter spacetime of interest.}. With this gauge fixation
freedom all metric elements are subject to nonvanishing
perturbations $H_{\ZA\ZB}(x,y)$ which can be decomposed into
$\ddim$-dimensional scalar, vector and tensor parts
$H_{\ZA\ZB}(x,y)=(H_{yy}(x,y),H_{y\alpha}(x,y),
H_{\alpha\beta}(x,y))$. The latter two in their turn can be
decomposed in irreducible scalar, transverse and
transverse-traceless components with respect to the fixed metric
$g_{\alpha\beta}(x)$ \footnote{Note, that throughout the paper
$\ddim$-dimensional (Greek) indices are lowered and raised with
the help of $g_{\za\zb}(x)$ and $\bnabla_{\za}$ denotes the
covariant derivative in the metric $g_{\za\zb}(x)$.}. Thus we have
    \begin{eqnarray}
    &&H_{yy}(x,y)=2\chi(x,y),            \label{lapse}\\
    &&H_{\za y}(x,y)=a^2(y)\big(v_\za(x,y)
    +\bnabla_\za b(x,y)\big),\qquad
    \bnabla^\zb v_\zb(x,y)=0,             \label{shift}\\
     &&H_{\za\zb}(x,y)=a^2(y)
     \big( \gamma_{\za\zb}(x,y)
     +2\bnabla_{(\za}\! f_{\zb)}(x,y)
     + 2\psi(x,y) g_{\za\zb}(x)
     + 2\bnabla_\za\!\bnabla_\zb\, e(x,y)\big),\nonumber\\
     &&\bnabla^\za\gamma_{\za\zb}(x,y)=
     g^{\za\zb}(x)\gamma_{\za\zb}(x,y)=0,\qquad
     \bnabla^\za f_\za(x,y)=0.      \label{Decomposition}
    \end{eqnarray}
Perturbations $H_{\za\zb}(x,y)$ are generated by the induced
metric perturbations $h^i=h_{\alpha\beta}^\pm(x)$
    \begin{eqnarray}
    &&h_{\alpha\beta}^\pm(x)=
    a^2(y_\pm)\Big(\gamma_{\alpha\beta}(x)
    + 2\psi(x) g_{\za\zb}(x)
    +2\bnabla_{(\za} f_{\zb)}(x)
    + 2\bnabla_\za\bnabla_\zb
    e(x)\Big)^\pm,        \label{Decomposition h}
   \end{eqnarray}
as boundary conditions for the linearized Einstein equations in
the bulk, $\psi(x,y_\pm)=\psi^\pm(x)$, $e(x,y_\pm)=e^\pm(x)$,
$f_\alpha(x,y_\pm)=f_\alpha^\pm(x)$,
$\gamma_{\za\zb}(x,y_\pm)=\gamma_{\za\zb}^\pm(x)$.

We know that the action (\ref{Action}) is invariant with respect
to $(d+1)$-dimensional diffeomorphisms.  Under their action with
the vector field $\xi^A(x,y)=(\xi^\alpha(x,y),\xi^y(x,y))$, whose
$d$-dimensional part has a transverse-longitudinal decomposition
in the metric $g_{\alpha\beta}(x)$
    \begin{eqnarray}
    \xi_\za(x,y)\equiv g_{\za\zb}(x)\xi^\zb(x,y)
    =\lambda_\za(x,y)+\bnabla_\za\mu(x,y),\qquad
    \bnabla_\za\lambda^\za=0,
    \end{eqnarray}
irreducible components of (\ref{lapse})-(\ref{Decomposition})
transform as
    \begin{eqnarray}
    &&\delta\chi=\ddy\xi^y, \\
    &&\delta b=a^{-2}\xi^y +\ddy \mu,\,\,\,
    \delta v_\alpha=\ddy \lambda_\za,\\
    &&\delta\psi=\psi+(\ddy \ln{a})\,\xi^y,
    \,\,\,\delta e=\mu,\,\,\,
    \delta f_\alpha(x,y)=\lambda_\za,\\
    &&\delta\gamma_{\za\zb}(x,y)=0.
    \end{eqnarray}
It is easy to check that the following combinations are invariant
under the action of these transformations \cite{HHR}
\newcommand{\A}{\mathcal{A}}
\newcommand{\B}{\mathcal{B}}
\newcommand{\V}{\mathcal{V}}
    \begin{eqnarray}
    &&\A=\ddy\psi+\psi\frac{a^{-2}R(g)}
    {\ddim(\ddim\!-\!1)\,\ddy\!\ln{a}}
    -\chi\,\ddy\ln{a}\,,\quad
    \B=\ddy e-b+\psi\frac{a^{-2}}
    {\ddy\ln{a}}\,,\quad          \nonumber\\
    &&\V_\za=\ddy f_\za-v_\za.   \label{invariants}
    \end{eqnarray}
The conformal part of the metric perturbation $\psi(x,y)$ is not
invariant under general diffeomorphisms, but for the class of
diffeomorphisms leaving the boundaries at constant values of the
$y$-coordinate, $\xi^y(x,y_\pm)=0$, boundary values $\psi^\pm(x)$
are also not transformed.

One can check that the quadratic part of the action expresses in
terms of these invariants, as it should be for the diffeomorphism
invariant functional. Direct calculation shows that it decomposes
in the sum of decoupled contributions of scalar, vector and
graviton (transverse-traceless) sectors
    \begin{eqnarray}
     \mathbf{S}^\DDim_{(2)}[\,H\,]=
     S_{\mathrm{scal}}[\,\A,\B,\psi\,]+
     S_{\mathrm{vect}}[\,\V\,]
     +S_{\mathrm{grav}}[\,\gamma\,],\label{QuadrActionDecomposition}
    \end{eqnarray}
which read\footnote{Up to total derivatives in $x^\za$ directions.
Derivatives in $y$-direction are kept, since they induce terms on
branes.}
    \begin{eqnarray}
     &&S_{\mathrm{scal}}[\,\A,\B,\psi\,]=
     2(\ddim-1)\int\limits_{\M}\dV\,
     \left(\ddim\A^2+2\A\,\bBox\B-
     \frac{R(g)}{\ddim(\ddim-1)}\B\,\bBox\B\right)\nonumber\\
     &&\qquad\qquad\qquad\qquad\qquad\qquad
     +2(\ddim-1)\sum_i\int
     \limits_{\dM_i}\dS\,
     \frac{a^{\ddim-2}}{\Bnabla_{\un}\ln{a}}\;
     \psi\left(\bBox
     +\frac{R(g)}{\ddim-1}\right)\psi,    \label{scalarpart}    \\
    &&S_{\mathrm{vect}}[\,\V\,]=
     \int\limits_{\M}\dV\,
     \V_{\za}\left(\bBox
     +\frac{R(g)}{\ddim}\right)\V^{\,\za},   \label{vectorpart}\\
     &&S_{\mathrm{grav}}[\,\gamma\,]=
    \frac12 \int\limits_{\M}\dV\,
     \Big(\!-\ddy\gamma_{\za\zb}\,
    \ddy\gamma^{\za\zb}
     +a^{-2}\gamma_{\za\zb}\bBox\gamma^{\za\zb}
     -\frac2{\ddim(\ddim-1)}a^{-2}R(g)
    \gamma_{\za\zb}
    \gamma^{\za\zb}\Big).\qquad               \label{gravitonpart}
    \end{eqnarray}
Here
      \begin{eqnarray}
        \bBox=g^{\za\zb}(x)\!
        \bnabla_{\!\za}\!\bnabla_{\!\zb}   \label{DefDalambertian}
      \end{eqnarray}
is the $\ddim$-dimensional covariant d'Alembertian. Note that it
is defined with respect to the $y$-independent metric
$g_{\za\zb}(x)$ rather than the background metric
$a^2(y)g_{\za\zb}(x)$. Also $\dV$ is the covariant Riemannian
measure in the bulk and $\dS$ is the covariant measure on branes
the latter again defined with respect to the auxiliary metric
$g_{\za\zb}(x)$
        \begin{eqnarray}
        &&\dV\equiv \dx\,dy\,
        a^\ddim(y)\sqrt{g(x)},          \label{MeasureBulk}\\
        &&\dS\equiv\dx\,\sqrt{g(x)}.    \label{MeasureBrane}
        \end{eqnarray}

\subsection{Scalar and vector sectors}
\hspace{\parindent}Interestingly, the calculation of scalar and
vector contributions in (\ref{QuadrActionDecomposition}) does not
require explicit solution of the bulk equations of motion. The
reason is that the invariants $\A$, $\B$ and $\V_\alpha$ enter the
Lagrangian without $y$-derivatives and, therefore, the same
property holds for the variables $\chi$, $b$ and $v_\za$
algebraically entering (\ref{invariants}). This in turn follows
from the fact that these variables play the role of Lagrange
multipliers of the nondynamical (in the extra ``time'' $y$)
$yA$-components of Einstein equations. So, varying the action with
respect to these Lagrange multipliers we have
    \begin{eqnarray}
     &&\frac{\delta \mathbf{S}^\DDim_{(2)}}{\delta \chi}=
     -(\ddy\ln{a})\,\frac{\delta \mathbf{S}^\DDim_{(2)}}{\delta
     \A}=-4a^\ddim\sqrt{g}\,
     (\ddim\!-\!1)(\ddy\ln{a})\,
     \big(\bBox\B+\ddim\!\cdot\!\A\big)=0,\nonumber\\
     &&\frac{\delta \mathbf{S}^\DDim_{(2)}}{\delta b}=
     -\frac{\delta \mathbf{S}^\DDim_{(2)}}{\delta
     \B}=-4a^\ddim\sqrt{g}\,(\ddim\!-\!1)\,
     \bBox\Big(\A-\frac{R(g)}
     {\ddim(\ddim\!-\!1)}\B\Big)=0,\nonumber\\
     &&\frac{\delta \mathbf{S}^\DDim_{(2)}}{\delta v_\za}=
     -\frac{\delta \mathbf{S}^\DDim_{(2)}}{\delta
     \V_\za}=-2a^\ddim\sqrt{g}\,
     \left(\bBox+\frac{R(g)}
     {\ddim}\right)\V^\za=0.                 \label{Constraints}
    \end{eqnarray}
In virtue of these equations the vector (\ref{vectorpart}) and the
{\em bulk} part of the scalar (\ref{scalarpart}) contributions to
the action vanish\footnote{Up to possible surface terms at the
infinity $x^\alpha$-coordinates in the bulk which we assume
vanishing because of vanishing sources everywhere except branes.}
and the result reduces to the brane term of (\ref{scalarpart}).

Note that the scalar sector is completely local and of second
order in derivatives. However, generally its kinetic terms are not
positive definite. With the factors $\Bnabla_{\un}\ln{a}$
expressed in terms of tensions (\ref{TensionsGen}) it takes the
form
    \begin{eqnarray}
    S_{\mathrm{scal}}[\,\psi^\pm\,]=
    -8(\ddim-1)^2\sum_{i=\pm}\int
     \limits_{\dM_i}\dS\,
     \frac{a^{\ddim-2}_i}{\sigma_i}\;
     \psi\left(\bBox
     +\frac{R(g)}{\ddim-1}\right)\psi,    \label{scalaraction}
    \end{eqnarray}
which clearly emphasizes its indefiniteness for two tensions of
opposite signs, $\sigma_+>0$, $\sigma_-<0$. On the positive
tension brane the conformal mode enters with the negative (ghost
like) sign -- exactly as in the Einstein theory\footnote{This does
not lead to physical instability because the conformal mode is not
dynamically independent in view of constraints.}, while on the
negative tension brane its sign is opposite. Only for
Anti-deSitter branes with $y_+<0<y_-$ both brane tensions can be
positive -- the case which we will not consider in detail.

For flat 4-dimensional branes, $R(g)=0$, $d=4$, this expression
reduces to the result obtained in \cite{BWEA}\footnote{To compare
(\ref{scalaraction}) with the equation (139) of \cite{BWEA} one
should bear in mind that our conformal modes $\psi^\pm$ are
rescaled relative to the conformal modes $\varphi^\pm$ in
\cite{BWEA}, $\varphi^\pm=2a_\pm^2\psi^\pm$.}.

\subsection{Graviton sector}
\hspace{\parindent}It is important that the tensor
(transverse-traceless) sector decouples from the rest of variables
not only in the action, but already at the level of solving the
boundary value problem -- $\gamma_{\alpha\beta}^\pm(x)$ (and only
$\gamma_{\alpha\beta}^\pm(x)$) serve as boundary conditions for
$\gamma_{\alpha\beta}(x,y)$. That is why the situation in this
sector literally repeats the technique of Sect.2. The action in
the graviton sector (\ref{gravitonpart}) after integration by
parts reads
   \begin{eqnarray}
     &&S_{\mathrm{grav}}[\,\gamma\,]=
    \frac12 \int\limits_{\M}\!\!\dV\,
     \gamma_{\za\zb}(\BBox\,
    \gamma^{\za\zb})+\frac12\,\sum_i\!\!\int
     \limits_{\dM_i}\!\!\dS\,a^\ddim\,
     \gamma_{\za\zb}
    \Bnabla_{\!\un}\gamma^{\za\zb},        \label{QuadrAction}
    \end{eqnarray}
where
    \begin{eqnarray}
     \BBox=a^{-\ddim}\ddy a^{\ddim}\ddy +
     a^{-2}\left(\bBox
    -\frac2{\ddim(\ddim-1)}R(g)\right)     \label{DefOperator}
    \end{eqnarray}
is a second order differential operator that governs the dynamics
of gravitons in the bulk. Comparison of the equation
(\ref{QuadrAction}) with (\ref{Wronskian2}) shows that the
Wronskian operator $\hat W_\pm$ coincides with $a^d_\pm\Bnablan$
on respective branes. Therefore, the solution of the Dirichlet
problem for transverse-traceless perturbations
    \begin{eqnarray}
     \left\{ \begin{array}{l}
     \BBox \gamma_{\za\zb}(x,y)=0\;; \\
     \\
     \gamma_{\za\zb} (x,y)|_{_{\dM_\pm}}
     =\gamma^\pm_{\za\zb}(x).
     \end{array} \right.                  \label{Dbvp for grav}
    \end{eqnarray}
in terms of the Dirichlet Green's function of the operator
(\ref{DefOperator})
    \begin{eqnarray}
     \left\{ \begin{array}{l}
     \BBox \GrD(x,y|x',y')=\delta(x,y|x',y')\;; \\
    \\
     \GrD(x,y|x',y')|_{_{\dM_\pm}}=0.
     \end{array} \right.                       \label{Dbvp for GF}
    \end{eqnarray}
where $\delta^\DDim (x,y|x',y')=
\delta^\ddim(x\!-\!x')\delta(y\!-\!y')/a^\ddim(y)\sqrt{g(x)}$ is
the covariant $\DDim$-dimensional $\delta$-function\footnote{Note
that \if{in contrast with Sect.2 we defined the operator
(\ref{DefOperator}) without the $\sqrt{G(x,y)}=a^d(y)\sqrt{g(x)}$
density weight, but in the Green's function definition (\ref{Dbvp
for GF}) the delta function is dedensitized by the same factor, so
that}\fi the kernel of the Green's function remains a biscalar
with respect to both of its arguments.}, reads as
    \begin{eqnarray}
     \gamma(x,y)=-\sum_{i=\pm}\int
     \limits_{\dM}\dS'\,\GrD(x,y|x',y_i)
    {\overleftarrow{W}}_i\gamma^i(x').
    \end{eqnarray}
Substituting it in (\ref{QuadrAction}) similarly to
(\ref{DActionFinal0}) results in the quadratic action in the
graviton
        \begin{eqnarray}
     &&S_{\mathrm{grav}}[\,\gamma\,]=
    -\frac12 \sum_{i,j}\int
     \limits_{\dM_i}\dS
    \int \limits_{\dM_j}\dS'\,
     \gamma^i(x)[\NGrDN]_{ij}(x,x')
     \gamma^j(x'),                             \label{DEffAction}
    \end{eqnarray}
where
    \begin{eqnarray}
    [\nGrDn]_{ij}(x,x')=
    a^\ddim_i \,a^\ddim_j\,\Bnabla_{\un_i}
    \Bnabla_{\un_j} \GrD (x,y_i|x',y_j)        \label{Dformfactor}
    \end{eqnarray}
and the primed integration measure is obviously
$\dS'=d^dx'\sqrt{g(x')}$.

In what follows the operator notations for the kernels of integral
operations will be more convenient for our purposes. This operator
form follows from the fact that the differential operator
(\ref{DefOperator}) depends on $\bBox$ as a parameter, therefore
the kernel of its inverse can be written down as a function of the
$\bBox$ acting on $\ddim$-dimensional $\delta$-function
$\delta^{(\ddim)}(x,x')\equiv\delta^\ddim(x\!-\!x')/\sqrt{g(x)}$
    \begin{eqnarray}
     [{\RnGrDnR}]_{ij}(\bBox)\,
     \delta^{(\ddim)}(x,x')
     =[{\RnGrDnR}]_{ij}(x,x').  \label{Kernel->Box}
    \end{eqnarray}

Thus, finally, after combining together tensor and scalar parts
and writing explicitly the integration measure, the quadratic part
of braneworld effective action looks as follows
    \begin{eqnarray}
     S^{(2)}_\ddim[\,h\,]= \frac12\int
     \limits_{\dM}\!\dx \sqrt{g(x)}
     \sum_{i,j=\pm}\!\left(-\gamma^i(x)
     [{\RnGrDnR}]_{ij}(\bBox)\,
     \gamma^j(x)+\psi^i(x)
     \mathrm{K}_{ij}(\bBox)
     \psi^j(x)\right),               \label{DActionFinal}
    \end{eqnarray}
where the notation $\psi^i(x)$ stands for $\psi(x,y_i)$ and
    \begin{eqnarray}
     \mathrm{K}_{ij}(\bBox)=
     4(\ddim-1)\frac{a_i^{\ddim-2}}
     {\Bnabla_{\un_i}\!\ln{a}}\,
     \delta_{ij}\left(\bBox+\frac{R(g)}
     {(\ddim-1)}\right).             \label{DilatonMatrix}
    \end{eqnarray}
This matrix operator for the case of $\ddim=4$ flat branes
coincides via appropriate rescaling with the operator ${\bf
K}_\Phi(\bBox)$ calculated in \cite{BWEA} by a different
procedure.

At this point it is worth discussing why the sector of
transverse-traceless modes in (\ref{DActionFinal}) can be regarded
as the graviton sector even for curved branes. The recovery of a
particle interpretation for the effective action with the
essentially nonlocal form factor (\ref{Kernel->Box}) is a matter
of the low-energy approximation -- the expansion of this form
factor in powers of $\bBox\to 0$. In this paper we will not
consider this expansion in much detail because it was
systematically developed in \cite{BWEA} for flat branes. Here we
will only sketch a similar procedure and its peculiarities due to
curvature of a homogeneous brane background.

To begin with, note that the $d$-dimensional term
$\bBox-2R(g)/\ddim(\ddim-1)$ of the operator $\BBox$,
(\ref{DefOperator}), also enters as a whole the quadratic term
$\gamma_{\za\zb}\big(\bBox-2R(g)/\ddim(\ddim-1)\big)\gamma^{\za\zb}$
of the $\gamma$-perturbation expansion for the $\ddim$-dimensional
Einstein-Hilbert action
    \begin{equation}
    S_{EH}[\,g\!+\!\gamma\,]=
    \int d^{\ddim}x\,
    \sqrt{\det(\,g\!+\!\gamma\,)}\,
    \big(R(\,g\!+\!\gamma\,)-2\lambda\big)
    \end{equation}
having a particular value of the cosmological constant
    \begin{equation}
    \lambda=\frac{\ddim-2}{2\ddim}\,R(g),
    \end{equation}
which supports the vacuum homogeneous background with a constant
scalar curvature $R(g)$. This is the reason why in the low-energy
approximation the usual Einstein gravity theory is recovered on a
positive tension brane even when it is not flat. The mechanism of
this phenomenon is the following. The nonlocal operator
(\ref{Kernel->Box}) of the transverse traceless sector
parametrically depends on $\bBox-2R(g)/\ddim(\ddim-1)$ as a whole
and, therefore, can be expanded in powers of this combination. One
can show that this expansion starts with the first order term (for
properly defined graviton fields diagonalising the quadratic form
in (\ref{DActionFinal}), which means that this mode is massless
\cite{BWEA,BWEA2}. The masslessness implies that on the maximally
symmetric spacetime,
$R_{\alpha\beta\mu\nu}(g)=
(g_{\alpha\mu}g_{\beta\nu}-g_{\alpha\nu}g_{\beta\mu})R(g)/d(d-1)$,
the solutions of the linearized equations of motion for {\em
transverse-traceless} modes
    \begin{equation}
    \left(\bBox-\frac{2R(g)}
    {\ddim(\ddim-1)}\right)
    \gamma_{\alpha\beta}(x)=0      \label{graveq}
    \end{equation}
still have a residual gauge freedom
$\gamma_{\za\zb}(x)\rightarrow\gamma_{\za\zb}(x)
+2\bnabla_{(\za}\xi_{\zb)}(x)$ with the transverse vector field,
$\bnabla_\za\xi^\za=0$, satisfying the equation
$2\bnabla^\zb\bnabla_{(\za}\xi_{\zb)}= (\bBox+R(g)/\ddim)\xi_\za
=0$. The transversality together with this equation guarantee that
these transformations preserve transverse-traceless nature of
$\gamma_{\alpha\beta}(x)$ and the equations of motion
(\ref{graveq}) because
$\big(\bBox-2R(g)/\ddim(\ddim-1)\big)\bnabla_{(\za}\xi_{\zb)}=0$.
Thus, these residual transformations reduce the number of physical
degrees of freedom for $\gamma_{\alpha\beta}$ by $d-1$ to that of
the $d$-dimensional massless graviton $d(d-3)/2$. This justifies
the interpretation of the transverse-traceless modes as gravitons.

\section{Alternative derivation of the braneworld effective
action} \label{Neumann}
\hspace{\parindent}There exists an
alternative derivation of the effective action suggested in
\cite{brane,BWEA}. It is based on the recovery of the action from
the effective equations for brane $d$-dimensional field. These
effective equations actually represent the Israel junction
conditions on branes rewritten entirely in terms of the brane
metric itself. In order to see this let us write down the metric
variation of the fundamental $\DDim$-dimensional action
(\ref{Action})
     \begin{eqnarray}
      &&\delta{\mathbf S}[\,G,\varphi\,]=
      -\int\limits_{\M\times Z_2} d^{d+1}X\,
      \sqrt{G}\left(\!R^{AB}(G)
      -\frac12\,\!R(G)\,
      G^{AB}+\Lambda G^{AB}\right)\delta G_{AB}(X)+  \nonumber\\
      &&\qquad\quad
      +\sum\limits_i\int\limits_{\dM_i}\!
      d^{\ddim}x\,
      \sqrt{g}\left([K^{\mu\nu}
      -g^{\mu\nu}K]
      +\frac12(T^{\mu\nu}
      -g^{\mu\nu}\tens_i)\right)
      \delta g_{\mu\nu}(x).            \label{Variation}
     \end{eqnarray}
Here $g_{\mu\nu}(x)$ denotes the induced metric on branes (in
contrast with the notations of the previous section, where
$g_{\mu\nu}(x)$ was reserved for the auxiliary homogeneous
metric), $\left[K^{\mu\nu}-g^{\mu\nu}K\right]$ is the jump of the
extrinsic curvature terms across the brane, and $T^{\mu\nu}(x)$ is
the corresponding $\ddim$-dimensional stress tensor of matter
fields on branes
    \begin{eqnarray}
     T^{\mu\nu}(x)=\frac2{\sqrt{g}}
     \frac{\delta S^{\rm mat}[g,\varphi]}
     {\delta g_{\mu\nu}(x)}.         \label{StressTensor}
    \end{eqnarray}
Note that the brane part of the (\ref{Variation}) gives the Israel
junction conditions.

Suppose that we solved Einstein equations of motion in the bulk
subject to fixed brane metrics and substituted the obtained
solution $G_{AB}=G_{AB}[\,g\,]$ in the original action in order to
find the effective one, $S^{\rm eff}[\,g,\varphi\,]={\mathbf
S}[\,G[\,g\,],\varphi\,]$. Then the variation of the latter will
not contain the bulk term of (\ref{Variation}) -- because the bulk
equations are satisfied -- and the metric variation of $S^{\rm
eff}[\,g,\varphi\,]$ will be
    \begin{eqnarray}
     \frac{\delta S^{\rm eff}[\,g,\varphi\,]}
     {\delta g_{\mu\nu}(x)}=
     \sqrt{g}\left([K^{\mu\nu}-g^{\mu\nu}K]
     +\frac12(T^{\mu\nu}
     -g^{\mu\nu}\sigma_i)
     \right)_{\dM_i}.       \label{BoundaryVariation}
    \end{eqnarray}
Here, certainly, the extrinsic curvatures containing the
derivatives normal to brane surfaces are the functions of brane
metrics. Thus, this equation can be functionally integrated to
recover $S^{\rm eff}[\,g,\varphi\,]$. The result
(\ref{BoundaryVariation}) is exact, but expressing its right hand
side in terms of brane metrics can in general be obtained only
within perturbation theory. This procedure was implemented in
\cite{BWEA} in the quadratic approximation in metric perturbations
on a flat brane background. To simplify the presentation we
briefly repeat this procedure in our case.

We start with the linearized version of the bulk Einstein
equations and linearized Israel matching conditions on the branes.
For transverse-traceless perturbations $\gamma_{\alpha\beta}(x,y)$
they reduce to the linear Neumann boundary value problem
    \begin{eqnarray}
     \left\{ \begin{array}{l}
     \BBox \gamma_{\za\zb}(x,y)=0,  \\
     \\
     a^\ddim\Bnablan\,\gamma_{\za\zb}
     (x,y)|_{_{\dM_i}}=
     -\frac12 t_{\alpha\beta}^i(x).\\
     \end{array} \right.                    \label{Nbvp for grav}
    \end{eqnarray}
On account of the constraints (\ref{Constraints}) in the bulk, the
vector part of the equations is trivially satisfied, while their
scalar part is locally restricted to branes and reads
\cite{GarTan}
    \begin{eqnarray}
     4(\ddim-1)\frac{a_i^{-2}}{\Bnabla_{\un_i}\ln{a}}
     \left(\bBox+\frac{R(g)}
     {(\ddim\!-\!1)}\right)\psi^i(x)=
     -t^i_{\mathrm{scal}}(x).              \label{Eq for dilaton}
    \end{eqnarray}
From now on we go back to notations of the previous section:
$\BBox$ is the operator (\ref{DefOperator}) acting in the bulk,
$\Bnablan$ is the derivative normal to the brane
(\ref{Def_nabla_n}), $t_{\za\zb}^{\,i}(x)$ is the {\em transverse
traceless} part of the rescaled matter stress tensor
$a^{\ddim-2}T_{\za\zb}(x)$ on the $i$-th brane and
$t^i_{\mathrm{scal}}= a^{\ddim-2}g^{\za\zb}(x)T_{\za\zb}^i(x)$
(where indices of $T_{\alpha\beta}^i(x)$ were lowered with the
induced metric on appropriate brane). Greek indices are raised and
lowered with the aid of the auxiliary fixed metric $g_{\za\zb}(x)$
of the homogeneous spacetime, which is conformally related to
background metrics of all branes (differing only by conformal
factors $a_i^2$) which agrees with convention of previous section.

The general solution of Neumann boundary value problem (\ref{Nbvp
for grav}) is\footnote{As in the case of the Dirichlet boundary
value problem above in what follows we drop for brevity Greek
indices in linear and quadratic combinations of
$\gamma_{\alpha\beta}(x,y)$ and $t_{\alpha\beta}^{\,i}(x)$.}
    \begin{eqnarray}
     \gamma(x,y)=-\frac12 \sum_i\int
     \limits_{\dM}\dS'\,\,
    \GrN(x,y|x'\!,y_i)\,t_i(x'),      \label{GenSolNbvp}
    \end{eqnarray}
where $\GrN(x,y,x'\!,y_i)$ is the Neumann Green's function
    \begin{eqnarray}
     \left\{ \begin{array}{l}
     \BBox \GrN(x,y|x',y')=\delta^{\DDim}(x,y|x',y'),  \\
    \\
     \Bnablan\GrN(x,y|x',y')|_{_{\dM_i}}=0.
     \end{array} \right.                      \label{Nbvp for GF}
    \end{eqnarray}

Effective dynamics of the brane metric allows one to express the
latter solely in terms of brane matter sources. In the
transverse-traceless sector this reduces to restricting the
general bulk solution to branes, i.e. by putting in
(\ref{GenSolNbvp}) $y=y_j$. In terms of the operator notation for
the kernel of the Neumann Green's function (similar to
(\ref{Kernel->Box})),
    \begin{eqnarray}
     \GrN(x,y_j|x',y_i)=
     {\mathbf G}^{j\,i}_N(\bBox)\,
     \delta^{(\ddim)}(x,x'),        \label{GenRestrSolNbvp}
    \end{eqnarray}
this reads as
    \begin{eqnarray}
     \gamma^j(x)=-\frac12 \sum_i
     {\mathbf G}^{j\,i}_N(\bBox)
     \,t_i(x).                    \label{NGravOperRescDynamics}
    \end{eqnarray}

There are infinitely many actions that generate by variational
procedure the last equation. Fortunately, there is a hint coming
from Eq.(\ref{BoundaryVariation}) that allows one to fix this
ambiguity -- the rescaled matter stress tensor enters the
variational derivative of the braneworld action with respect to
induced metric $a^2 g_{\alpha\beta}(x)$  with the algebraic
coefficient $a^{-2}\sqrt{g(x)}$. Therefore, by rewriting
(\ref{NGravOperRescDynamics}) and (\ref{Eq for dilaton}) as
    \begin{eqnarray}
     &&\sqrt{g(x)}\sum_j
     \left({\mathbf G}^{-1}_N\right)_{ij}(\bBox)\,
     \gamma^j(x)+\frac12\sqrt{g(x)}\,t_i(x)=0,     \nonumber\\
     &&\sqrt{g(x)}\,\sum_j
     \mathrm{K}_{ij}(\bBox)\psi^j(x)+
     \sqrt{g(x)}\, t_i^{\mathrm{scal}}(x)=0,       \label{NEqns}
    \end{eqnarray}
where $\mathrm{K}_{ij}(\bBox)$ is defined by
(\ref{DilatonMatrix}), and $\left({\mathbf
G}^{-1}_N\right)_{ij}\!(\bBox)$ is the inverse of ${\mathbf
G}^{j\,i}_N(\bBox)$
    \begin{eqnarray}
     \sum_j\left({\mathbf G}^{-1}_N\right)_{ij}\!(\bBox)
     {\mathbf G}^{j\,k}_N(\bBox)=\delta^k_i,
    \end{eqnarray}
one easily finds the action generating (\ref{NEqns})
    \begin{eqnarray}
     &&S^{\rm eff}[\,h,\varphi\,]= \frac12\int
     \limits_{\dM}\!\dx \sqrt{g(x)}\sum_{i,j}\left(
     \gamma^i(x)
     \left({\mathbf G}^{-1}_N\right)_{ij}\!(\bBox)\,
     \gamma^j(x)+\psi^i(x) \mathrm{K}_{ij}(\bBox)
     \psi^j(x) \right)                          \nonumber\\
     &&\qquad\qquad\qquad\qquad+
     \frac12\int\limits_{\dM}\!\dx \sqrt{g(x)}\,
     \sum_i\left(\gamma^i(x)t_i(x)
     +2\psi^i(x)
     t_i^{\mathrm{scal}}(x)\right).      \label{NActionFinal}
    \end{eqnarray}
The source term with $t^i(x)$ and its trace here is obviously a
linear in metric perturbation part of the action of matter field,
while the quadratic in $\gamma$ part has a purely gravitational
nature and, thus, should necessarily coincide with the action
(\ref{DActionFinal}) obtained within the Dirichlet boundary
conditions. This implies the equality of two operators arising in
two different boundary value problems -- the Dirichlet and Neumann
one
    \begin{eqnarray}
     [{\rnGrDnr}]_{ij}(\bBox)= -
    \left({\mathbf G}^{-1}_N\right)_{ij}(\bBox). \label{Relation}
    \end{eqnarray}

\section{Duality of boundary value problems}
\renewcommand{\M}{{\cal M}}
\renewcommand{\dM}{{\partial \cal M}}
\hspace{\parindent}To prove the duality relation (\ref{Relation})
let us, first, remind the notations of Sect.2. We consider $\M$ --
the bulk spacetime with branes/boundaries $\dM_i$:
$\dM=\bigcup_i\dM_i$. Let $X$ be the coordinates on the bulk $\M$
and $x$ -- the coordinates on the branes $\dM_i$, so that the
embedding of $\dM_i$ into $\M$ reads as $X=X_i(x)$. Introduce $H$
-- the field in the bulk and denote boundary values of the bulk
field on $\dM_i$ by $h^i$, $h^i(x)=H(X_i(x))$. Let $\BBox$ be some
nondegenerate self-adjoint differential operator of the second
order in derivatives acting on $H(X)$. Let it satisfy the
Wronskian relation (\ref{Wronskian}) with the first order
Wronskian operator $\hat W$.

Let $H(X)$ satisfy the Neumann problem with the inhomogeneous
boundary conditions with some given sources $j_k(x)$ on the
boundary
    \begin{eqnarray}
     \left\{ \begin{array}{l}
     \BBox\,H(X)=0\;; \\
      \\
     \hat W \,H(X) |_{X=X_k(x)}=j_k(x).
     \end{array} \right.                      \label{Nbvp}
    \end{eqnarray}
The solution of this problem in terms of the Neumann Green's
function of the operator $\BBox$, satisfying\footnote{We define
the delta-function $\delta(X,X')$ by the relation
$\int\limits_{\M} \dV H(X)\,\delta(X,X')=H(X')$.}
    \begin{eqnarray}
     \left\{ \begin{array}{l}
     \BBox \,\GrN(X,X')=\delta(X,X'), \\
    \\
     \hat W \,\GrN(X,X')|_{_{X\in\dM_k}}=0,
     \end{array} \right.                    \label{NGfbvp}
    \end{eqnarray}
reads
    \begin{eqnarray}
     H(X)=\sum_k\,\int
     \limits_{\dM_k}\dS'\,
    \GrN\big(X,X_k(x')\big)j_k(x').  \label{NGenSol}
    \end{eqnarray}

Its restriction to the boundary yields the relation between the
boundary fields $h^l(x)$ and their conjugates -- sources $j_k(x)$
        \begin{eqnarray}
         h^l(x)=\sum_k\,
         \int\limits_{\dM_k}\dS'\,
         {\mathbf G}^{lk}_N(x,x')j_k(x')   \label{N}
        \end{eqnarray}
in terms of the new boundary operator kernel
        \begin{eqnarray}
        {\mathbf G}^{lk}_N(x,x')=
         \GrN\big(X_l(x),X_k(x')\big)   \label{N kernel}
        \end{eqnarray}

On the other hand the same field (\ref{NGenSol}) serves as the
solution of the Dirichlet problem (\ref{dF(df)}) with the boundary
data $H(X)|_{X=X(x)}=h(x)$ related to the sources $j(x)$ according
to (\ref{N}). Therefore, in terms of the Dirichlet Green's
function  (\ref{DGfbvp}) $H(X)$ has the form (\ref{DGenSol}).
Acting on (\ref{DGenSol}) on the boundaries by the Wronskian and
taking into account that $\hat W H(X)|_{_{\dM_l}}\!=j_l(x)$ one
finally obtains the alternative to (\ref{N}) relation between
$h(x)$ and $j(x)$
    \begin{eqnarray}
     j_k(x)=-\sum_l\,\int \limits_{\dM_l}\dS'\,
     [\overrightarrow{W}\GrD
    \overleftarrow{W}]_{kl}(x,x')\,h^l(x'),       \label{D}
    \end{eqnarray}
where the kernel
$[\overrightarrow{W}\GrD\overleftarrow{W}]_{kl}(x,x')$ is defined
by Eq.(\ref{matrix}).

Comparing (\ref{N}) and (\ref{D}) one concludes that ${\mathbf
G}^{lk}_N(x,x')$ and
$-[\overrightarrow{W}\GrD\overleftarrow{W}]_{kl}(x,x')$ are the
kernels of the inverse operations on $\dM$
    \begin{eqnarray}
    -\sum_i\,\int \limits_{\dM_i}\dS'\,
     [\overrightarrow{W}\GrD
    \overleftarrow{W}]_{li}(x,x'')\,
    {\mathbf G}^{ik}_N(x,x')
    =\delta^l_k\,\delta(x,x').       \label{ND}
    \end{eqnarray}
This confirms the operator relation (\ref{Relation}). Below we
consider two examples demonstrating the efficiency of this result.

\subsection{Simple example}
\hspace{\parindent}Consider the simplest model with the operator
on the one-dimensional space of the coordinate $y$, $y_1\leq y\leq
y_2$,
    \begin{eqnarray}
     \BBox=\frac{\partial^2}
     {\partial y^2} + m^2,          \label{1dOperator}
    \end{eqnarray}
and introduce its Wronskian operator acting on boundaries
    \begin{eqnarray}
     &&\Bnablan =
     \left\{\begin{array}{l}
     +\ddy, \qquad y=y_1, \\
     -\ddy, \qquad y=y_2.
     \end{array}\right.
    \end{eqnarray}
The Dirichlet and Neumann Green's functions of this operator read
    \begin{eqnarray}
     &&\GrD(y,y')=
     \frac{\sin{m(y'\!-\!y_2)}\sin{m(y\!-\!y_1)}}
     {m\sin{m(y_2\!-\!y_1)}}\theta(y'\!-\!y)
     +(y\leftrightarrow y'),\\
     &&\GrN(y,y')=
     \frac{\cos{m(y'\!-\!y_2)}\cos{m(y\!-\!y_1)}}
     {m\sin{m(y_2\!-\!y_1)}}\theta(y'\!-\!y)
     +(y\leftrightarrow y'),
    \end{eqnarray}
so that the operators (\ref{matrix}) and (\ref{N kernel}) turn out
to be $2\times2$-matrices
    \begin{eqnarray}
     &&-[\overrightarrow{\Bnabla}_\un
     \GrD\overleftarrow{\Bnabla}_\un]_{ij}=
     \frac{m}{\sin{m(y_2\!-\!y_1)}}
     \left(
     \begin{array}{cc}
      \!\!-\cos{m(y_2\!-\!y_1)}
      &\! 1 \\
      &\!\!\\
      \!\! 1
      &\!\!-\cos{m(y_2\!-\!y_1)}
     \end{array}\!\!\!\right),          \label{1d nGrDn}\\
     &&\GrN^{ij}=
     \frac{1}{m\sin{m(y_2\!-\!y_1)}}
     \left(
     \begin{array}{cc}
      \!\!\cos{m(y_2\!-\!y_1)}
      &\! 1 \\
      &\!\!\\
      \!\! 1
      &\!\!\cos{m(y_2\!-\!y_1)}
     \end{array}\!\!\!\right).          \label{1d GrN}
    \end{eqnarray}
This is a matter of direct verification that these matrices are
inverse to one another, because their determinants equal
    \begin{eqnarray}
     \det{\GrN}^{ij}=\Big(\det[\nGrDn]_{ij}\Big)^{-1}
     =-\frac{1}{m^2}.
    \end{eqnarray}

\subsection{Example: two-brane Randall-Sundrum model}
\newcommand {\w}{\mathbf{\varkappa}}
\newcommand {\wpp}{\varkappa_\p}
\newcommand {\wnn}{\varkappa_\n}
\hspace{\parindent}In this section we illustrate the relation
(\ref{Relation}) for the Randall-Sundrum model with two flat
branes. As in Sect.3 we work in coordinate system
$X^\ZA=(x^\za,y)$ in which background metric has the form
(\ref{Background}) with flat spacetime metric
$g_{\alpha\beta}(x)=\eta_{\alpha\beta}$. Two  flat branes with
tensions $\tens_\p=-\tens_\n$ are located respectively at $y=\yp$
and $y=\yn$. In such a coordinate system
    \begin{eqnarray}
     &&\BBox=a^{-\ddim}\ddy a^{\ddim}\ddy +
     a^{-2}\bBox\,;\qquad \bBox
    =\eta^{\za\zb}\bnabla_{\za}\bnabla_{\zb}\,,\\
     &&\eta^{\za\zb}(x)\eta_{\za\zb}(x)
    =\ddim\,,\qquad a(y)=\exp(-
     2ky)\,,\nonumber\\
     &&\Bnablan \equiv\un^\ZA\Bnabla_\ZA=
     \left\{\begin{array}{l}
     +\ddy\,, \qquad y=\yp, \\
     -\ddy\,, \qquad y=\yn.
     \end{array}\right.               \label{RSQuantities}
    \end{eqnarray}

For the sake of brevity we introduce the new operator-valued
variable depending on the coordinate $y$
    \begin{eqnarray}
     &&\w\equiv\w(y,\bBox)
    =\frac{\sqrt{\bBox}}{k\,a(y)}\;,\nonumber\\
     &&\wpp=\w(\yp,\bBox)\;,
    \quad\wnn=\w(\yn,\bBox)\;,          \label{Def w}
    \end{eqnarray}
and the following two-point combinations of Bessel functions
    \begin{eqnarray}
     &&\D(\w,\w')=\frac{\pi}{2}\left(
      Y_{\ddim/2}(\w)J_{\ddim/2}(\w')-
      Y_{\ddim/2}(\w')J_{\ddim/2}(\w)\right),\nonumber\\
     &&\E(\w,\w')=\frac{\pi}{2}\left(
      Y_{\ddim/2}(\w)J_{\ddim/2-1}(\w')-
      Y_{\ddim/2-1}(\w')J_{\ddim/2}(\w)\right),\nonumber\\
     &&\C(\w,\w')=\frac{\pi}{2}\left(
      Y_{\ddim/2-1}(\w)J_{\ddim/2-1}(\w')-
      Y_{\ddim/2-1}(\w')
    J_{\ddim/2-1}(\w) \right).          \label{DefDEC}
    \end{eqnarray}
Then the explicit expressions for Dirichlet and Neumann Green
functions take the form \cite{BWEA}
    \begin{eqnarray}
     &&\GrD(y,y'|\bBox)=\frac{\D(\w',\wnn)\D(\w,\wpp)\,
    \theta(y'\!-\!y)
     +\D(\w',\wpp)\D(\w,\wnn)\,\theta(y\!-\!y')}
     {a(y)^{\ddim/2}a(y')^{\ddim/2}\,k\,\D(\wnn,\wpp)}\;,\\
     &&\GrN(y,y'|\bBox)=\frac{\E(\w',\wnn)\E(\w,\wpp)\,
    \theta(y'\!-\!y)
     \,+\,\E(\w',\wpp)\E(\w,\wnn)\,\theta(y\!-\!y')}
     {a(y)^{\ddim/2}a(y')^{\ddim/2}\,k\,
    \C(\wnn,\wpp)}\;.
     \end{eqnarray}

After acting on the Dirichlet Green's function with normal
derivatives (\ref{RSQuantities}) and restricting them to branes
($y,y'=\yp,\yn$) one obtains the $2\times2$-matrix valued operators
-- the kernel of the braneworld effective action in the Dirichlet
setup (\ref{DActionFinal}),
    \begin{eqnarray}
     -[\RnGrDnR]_{ij}(\bBox)=
     \frac{k}{\D(\wnn,\wpp)}
     \left(
     \begin{array}{cc}
      \!\!a^\ddim_\p \wpp\E(\wnn,\wpp)
      &\! a^{\ddim/2}_\p a^{\ddim/2}_\n \\
      &\!\!\\
      \!\! a^{\ddim/2}_\p a^{\ddim/2}_\n
      &\!\!a^\ddim_\n\wnn\E(\wpp,\wnn)
     \end{array}\!\!\!\right),          \label{RS Operator}
    \end{eqnarray}
and the inverse of the kernel in the Neumann setting
(\ref{NActionFinal})
    \begin{eqnarray}
     {\mathbf G}^{ij}_N(\bBox)=
    \frac{\left(k\wpp\wnn\right)^{-1}}
     {\C(\wnn,\wpp)}
     \left(
     \begin{array}{cc}
      \!\!-a^{-\ddim}_\p \wnn\E(\wpp,\wnn)
      &\! a^{-\ddim/2}_\p a^{-\ddim/2}_\n \\
      &\!\!\\
      \!\! a^{-\ddim/2}_\p a^{-\ddim/2}_\n
      &\!\!-a^{-\ddim}_\n \wpp\E(\wnn,\wpp)
     \end{array}\!\!\!\right).         \label{RS InvOperator}
    \end{eqnarray}
Here we used the properties of Bessel functions
\cite{AbramowitzStegun}
    \begin{eqnarray}
     \partial_{\w}(\w^{\ddim/2}\D(\w',\w))=
     \w^{\ddim/2}\E(\w',\w)\,,
     \qquad\E(\w,\w)=-\w^{-1}\,.
    \end{eqnarray}

The calculation of the $2\times2$-determinants of these matrices
can be easily done by using a simple relation
    \begin{equation}
    \E(\w',\w')\E(\w,\w)-\E(\w',\w)\E(\w,\w')
    =\D(\w,\w')\C(\w,\w')
    \end{equation}
from which it follows that
    \begin{eqnarray}
     \det{\mathbf G}^{ij}_N(\bBox)=
     \Big(\det[\RnGrDnR]_{ij}(\bBox)\Big)^{-1}=
     -\frac{(k^2\wnn\wpp)^{-1}}{a^{\ddim}_\p \;a^{\ddim}_\n}
     \frac{\D(\wnn,\wpp)}{\C(\wnn,\wpp)}.
    \end{eqnarray}
Then it is a matter of direct check that the matrices (\ref{RS
Operator}) and (\ref{RS InvOperator}) are inverse to one another.

\section{Conclusions}
\hspace{\parindent}Thus we obtained the braneworld effective
action in the weak field (quadratic) approximation for models of
curved branes with the deSitter and Anti-deSitter geometries,
embedded in the Anti-deSitter bulk. The calculation was done
within two different schemes resorting to Dirichlet and Neumann
boundary value problems, and their equivalence was shown on the
basis of a special relation between the boundary operators
associated respectively with the Dirichlet and Neumann Green's
functions of the theory. Although this relation was used here only
in a braneworld context, it seems to have a much wider scope of
implications in various models relating volume (bulk) and surface
phenomena, like graviton localization \cite{RS}, AdS/CFT
correspondence and holography principle
\cite{BalKraLa,Gubser,holography,SkendSol,SachsSol}.

Regarding the braneworld action algorithms for curved branes --
this result can be important for the brane models of inflation
scenario \cite{Tye,brane,BWEA}. One of the motivations for
considering deSitter type branes is that they model the inflation
stage in the dynamics of the brane Universe embedded in the
multidimensional Anti-deSitter bulk. The idea of the radion field
playing the role of the inflaton was suggested in \cite{brane},
where inflation was induced by the effectively repulsive force
between the branes diverging in the course of inflationary stage.
This repulsive force arises in virtue of the detuning of brane
tensions from their Randall-Sundrum values (\ref{TensionsFlat})
corresponding to flat branes. However, the analysis of this
detuning in \cite{brane,BWEA} was done by directly extrapolating
the flat branes results to curved branes. Now, with the knowledge
of the answer (\ref{DActionFinal}) the effects of curvature can be
studied at a rigorous quantitative level. This and, in particular,
the effects of brane curvature on the low-energy limit of the
nonlocal form factor (\ref{Dformfactor})-(\ref{Kernel->Box}) will
be considered elsewhere.

Another possible generalization of the obtained results concerns
going beyond the limitation of conformally equivalent branes. Note
that everywhere in the constructions above both branes are
``parallel'' surfaces simultaneously of the flat, deSitter or
Anti-deSitter geometry. It would be interesting to generalize
calculations to coexisting branes of different types (different
signs of curvature), because this might lead to the dynamical
description of intersecting curved branes very popular in recent
years \cite{Reall,KOST,Pyr,Neronov,Bucher}. Finally, the algorithm
(\ref{DActionFinal}) suggests a natural generalization to
multibrane cases when the range of index $i$ goes beyond two
possible values. This generalization particularly implies that a
brane can serve as a boundary of more than two smooth bulks glued
to it. Such a construction of branching bulks can be shown to have
explicit Randall-Sundrum type solutions of Einstein equations with
appropriately generalized Israel junction conditions
\cite{Nesterov}. It forms a kind of spacetime bulk-brane foam
resembling selfreproducing inflationary Universe \cite{Linde} and
suggesting new interesting facets in the cosmological constant
problem.

\section*{Acknowledgements}
\hspace{\parindent}The work of A.O.B was supported by the RFBR
grant No 02-01-00930. The work of D.V.N. was supported by the RFBR
grant No 02-02-17054 and by the Landau Foundation. This work was
also supported by the RFBR grant No 00-15-96566.

\end{document}